\documentclass[%
 reprint,
onecolumn,
amsmath,amssymb,
 aps,
]{revtex4-1}

\usepackage{graphicx}% Include figure files
\usepackage{subcaption}
\usepackage{bm}% bold math
\usepackage{footnote}
\usepackage[utf8]{inputenc}
\usepackage{color, colortbl}
\usepackage{mhchem}
\usepackage{tabularx}
\usepackage{tabu}
\usepackage{braket}
\usepackage{comment}

\begin{document}

\title{Superfluid Condensate Fraction and Pairing Wave Function of the Unitary Fermi Gas}

\author{Rongzheng He}
\email{herongzh@msu.edu}
\author{Ning Li}
\email{lini@nscl.msu.edu}
\author{Bing-Nan Lu}
\email{lub@nscl.msu.edu}
\author{Dean Lee}
\email{leed@frib.msu.edu}
\affiliation{
Facility for Rare Isotope Beams and Department of Physics and Astronomy, Michigan State University, East Lansing, MI 48824, USA}

% \collaboration{MUSO Collaboration}%\noaffiliation

% \author{Charlie Author}
%  \homepage{http://www.Second.institution.edu/~Charlie.Author}
% \affiliation{
%  Second institution and/or address\\
%  This line break forced% with \\
% }%
% \affiliation{
%  Third institution, the second for Charlie Author
% }%
% \author{Delta Author}
% \affiliation{%
%  Authors' institution and/or address\\
%  This line break forced with \textbackslash\textbackslash
% }%

% \collaboration{CLEO Collaboration}%\noaffiliation

\date{\today}% It is always \today, today,
% \date{Nov }         %  but any date may be explicitly specified
\begin{abstract}
   The unitary Fermi gas is a many-body system of two-component fermions with zero-range interactions tuned to infinite scattering length.  Despite much activity and interest in unitary Fermi gases and its universal properties, there have been great difficulties in performing accurate calculations of the superfluid condensate fraction and pairing wave function.  In this work we present auxiliary-field lattice Monte Carlo simulations using a novel lattice interaction which accelerates the approach to the continuum limit, thereby allowing for robust calculations of these difficult observables.  As a benchmark test we compute the ground state energy of 33 spin-up and 33 spin-down particles.  As a fraction of the free Fermi gas energy $E_{FG}$, we find $E_0/E_{FG}= 0.369(2), 0.372(2),$ using two different definitions of the finite-system energy ratio, in agreement with the latest theoretical and experimental results.  We then determine the condensate fraction by measuring off-diagonal long-range order in the two-body density matrix.  We find that the fraction of condensed pairs is $\alpha = 0.43(2)$.  We also extract the pairing wave function and find the pair correlation length to be $\zeta_pk_F = 1.8(3) \hbar$, where $k_F$ is the Fermi momentum.  Provided that the simulations can be performed without severe sign oscillations, the methods we present here can be applied to superfluid neutron matter as well as more exotic P-wave and D-wave superfluids.
\end{abstract}
\maketitle

\section{Introduction}

The unitary Fermi gas describes an idealized limit of two-component fermions where the interactions have zero range and the scattering length is infinite.  The intense interest in the unitary limit reflects the fact that it describes universal physics that can be realized in ultracold atoms and also approximately describes neutron gases in the inner crust of a neutron star. It lies in a crossover region connecting a Bardeen-Cooper-Schrieffer (BCS) superfluid at weak coupling and Bose-Einstein condensate (BEC) at strong coupling \cite{paper 10, paper 11, paper 12}. The unitary limit is special in that the system has no intrinsic length scales.  Hence we can use simple dimensional analysis to determine the scaling of any observable as a function of the Fermi momentum. Hence the ground-state energy $E_0$ must proportional to the free Fermi gas energy $E_{FG}$,
\begin{equation}
E_0 = \xi E_{FG},
\end{equation}
where $\xi$ is a universal parameter, sometimes called the Bertsch parameter. $\xi$ has been measured by numerous experiments using the ultracold trapped atoms \cite{paper 4.32+1, paper 4.32+2, paper 4.32+3, paper 4.32+4, paper 4.32+5, paper 4.17, paper 4.32+7, paper 4.32+8, paper 4.41, paper 4.40, paper 4.32+11, paper 4.38} and also calculated by analytical methods \cite{paper 4.32+14, paper 4.32+15, paper 4.32+16, paper 4.32+17, paper 4.32+18,paper 4.32+19,paper 4.32+20,paper 4.32+21,paper 4.37,paper 4.32+23,paper 4.36,paper 4.32+25,paper 4.32+26,paper 4.32+27}. In addition, a substantial number of numerical calculations have been made using quantum Monte Carlo methods and other techniques \cite{paper 4.32+28, paper 4.32+29, paper 4.32+30, paper 4.32+31, paper 4.32+32,paper 4.32+33,paper 4.32+34,paper 4.32+35,paper 4.32+36,paper 4.34, paper 4.53, paper 4.35, paper 4.32+40, paper 4.32+41, paper 4.33,paper 4.52, paper 4.32, paper 4.32+42, paper 4.38+1, paper 4.38+2, Lee:2006vp}.

While there have been numerous calculations of the ground state energy, there have been no first principles calculations of the superfluid condensate fraction and pairing wave function using two-sided expectation values of the two-body density matrix.  The two-body density matrix, which is defined later in Eq.~(\ref{def_TBDM}), is extremely difficult to calculate in diffusion Monte Carlo calculations as it involves particle trajectories that have disconnected jumps.  Therefore previous investigations of the condensate fraction have computed one-sided expectation values, with the ground state on one side and an approximation to the ground state on the other.  In this letter we use auxiliary-field lattice Monte Carlo simulations and are able to eliminate this source of error in our calculations of the superfluid condensate fraction and pairing wave function.  The two-body density matrix calculated in this manner is very challenging due to large stochastic fluctuations \cite{Lee:2006vp}.  In the results presented here we are able to overcome these problems with high statistics and by using a new lattice action with faster convergence to the continuum limit.   

We express all quantities in units of the spatial lattice spacing $a_{\rm latt}$, particle mass $m$, and $\hbar$.  In these units the lattice time step used is $0.1065\,ma_{\rm latt}^2\hbar^{-1}$. Between particles with opposite spins we employ an attractive nonlocal S-wave interaction. By nonlocal we mean that the interaction is dependent on the velocity of the particles.  Nonlocal interactions have not previously been applied to first principles calculations of the unitary limit.  We find that it provides excellent control of the S-wave scattering parameters while zeroing out interactions in higher angular momentum channels.  We have adjusted the range of the interaction so that bulk properties like the ground state energy and condensate fraction converge as rapidly as possible in the dilute limit.  Our interaction parameters correspond to an infinite scattering length $a_s$, and an effective range of $r_e \sim 0.05$ lattice units.   As done in previous lattice calculations \cite{paper 4.32+32, paper 4.32+33, paper 4.32+34, paper 4.32+35, paper 4.34, paper 4.52, paper 4.53, paper 4.35, paper 4.32+40, paper 4.32+42, paper 4.32, paper 4.38+2, paper 4.38+1, Lee:2006vp}, we have decoupled the interactions between particles using an auxiliary field.  Details of the lattice Hamiltonian and the auxiliary-field Monte Carlo formalism are provided in the Supplemental Material.

We define the one-body density matrix (OBDM) as
\begin{equation}
    \rho_{1,\sigma}(\mathbf{r'},\mathbf{r}) = \langle a_{\sigma}^{\dagger}(\mathbf{r'})a_{\sigma}(\mathbf{r}) \rangle,
\end{equation}
and the two-body density matrix (TBDM) as
\begin{align}
   \rho_2(\mathbf{r'_1},\mathbf{r'_2}, & \mathbf{r_1},\mathbf{r_2})  
   = \nonumber \\
   & \braket{\Psi_0 |  a^{\dagger}_\uparrow(\mathbf{r'_1})a^{\dagger}_\downarrow(\mathbf{r'_2})
    a_\downarrow(\mathbf{r_2})
    a_\uparrow(\mathbf{r_1}) | \Psi_0}, \label{def_TBDM}
\end{align}
where $a_{\sigma}^{\dagger}(\mathbf{r})$, $a_{\sigma}(\mathbf{r})$ denote the creation and annihilation operators of a fermion at site $\mathbf{r}$ with spin $\sigma$.  Due to superfluid pairing, there are no long-range correlations in the OBDM, and so $\rho_{1,\sigma}(\mathbf{r'},\mathbf{r})$ vanishes as $|\mathbf{r} - \mathbf{r'}| \rightarrow \infty$ \cite{paper 4.43, paper 4.44}. In direct contrast, there will be long-range order in the TBDM, a signature of pairing in the superfluid phase of fermionic systems \cite{paper 13}. This is called off-diagonal long-range order and in quantitative terms means that $\rho_2(\mathbf{r'_1},\mathbf{r'_2},\mathbf{r_1},\mathbf{r_2})$ has an eigenvalue of the order of the total number of particles $N$ in the limit of large separation between the primed and unprimed coordinates. 
In the large separation limit, the TBDM factorizes as
\begin{equation}
    \rho_2(\mathbf{r'_1},\mathbf{r'_2},\mathbf{r_1},\mathbf{r_2}) = \alpha N/2 \cdot \phi^{\ast}(|\mathbf{r'_1} - \mathbf{r'_2}|) \phi(|\mathbf{r_1} - \mathbf{r_2}|),
    \label{TBDM}
\end{equation}
where $\phi(|\mathbf{r}|)$ is the normalized S-wave pair wave function and $\alpha$ is the condensate fraction, or the percentage of pairs with zero total momentum forming the superfluid condensate. Several experiments and theoretical calculations have been performed to determine $\alpha$ at unitarity, but this value has not yet been fully settled. Experiments in ultracold $\ce{^{6}Li}$ have found the condensate fraction to be 0.46(7) \cite{paper 4.47, paper 4.48} and 0.47(7) \cite{Roati}.  On the other hand another measurement \cite{paper 4.48+1} observed that at most 15\% of atom pairs were condensed in $\ce{^{40}K}$. Based on the BCS equation of state, zero-temperature calculations \cite{paper 4.46, paper 4.49} have shown the condensate fraction to be around 0.7, while the result $\alpha = 0.43$ has been predicted in Ref.~\cite{paper 4.48+2}. From quantum Monte Carlo methods, $\alpha$ was estimated to be 0.57(2) \cite{paper 4.51} and 0.56(1) \cite{paper 4.52}, but a more recent quantum Monte Carlo calculation yielded $\alpha = 0.51$ \cite{paper 4.52+1}.

\section{Methods}
\subsection{Lattice interactions}

We express all quantities in units of the spatial lattice spacing $a_{\rm latt}$, particle mass $m$, and $\hbar$.  In these units the lattice time step is $0.1605ma_{\rm latt}^2\hbar^{-1}$.  For reasons of notational convenience, in the intermediate steps where we describe the details the lattice calculations, we use lattice units.  This corresponds to omitting writing factors of $a_{\rm latt}$ and $\hbar$.  Our lattice system consists of an $L^3$ cube with periodic boundary conditions. The notation $\sum_{\langle \mathbf{n}\;\mathbf{n'}\rangle}$ indicates summation over nearest-neighbor lattice sites adjacent to $\mathbf{n}$ while $\sum_{\langle \mathbf{n}\;\mathbf{n'}\rangle _i}$ represents summation over nearest-neighbor sites of $\mathbf{n}$ along $i^{th}$ spatial axis. Similarly, $\sum_{\langle \langle \mathbf{n}\;\mathbf{n'}\rangle \rangle _i}$ is the sum over second nearest-neighbor of $\mathbf{n}$ along $i^{th}$ axis and $\sum_{\langle \langle \langle \mathbf{n}\;\mathbf{n'}\rangle \rangle \rangle _i}$ is the sum over third nearest-neighbor of $\mathbf{n}$ along $i^{th}$ axis.

\begin{figure}[h!]
    \centering
    \includegraphics[width=0.9\textwidth]{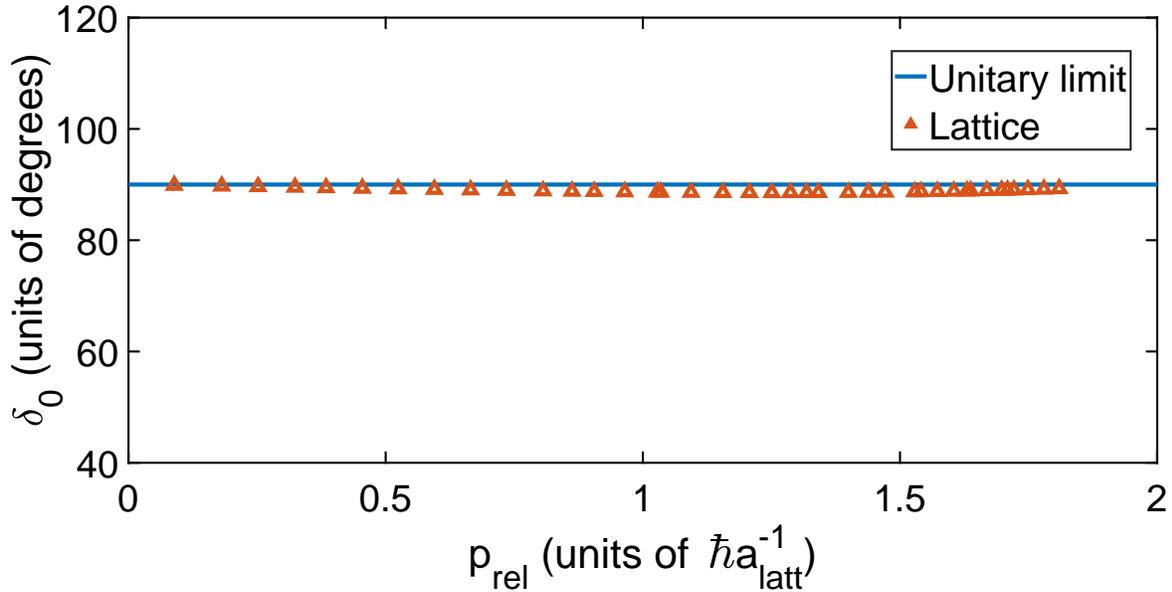}
    \captionsetup{justification=raggedright}
    \caption{Plot of the S-wave scattering phase shifts versus relative momentum.  Lattice results are shown with orange triangles and the unitary limit relation $\delta_0 = 90^{\circ}$ is indicated by the blue line.}
    \label{Fig. A.1}
\end{figure}

We use a free lattice Hamiltonian of the form
\begin{eqnarray}
\begin{split}
    H_{\rm free} &= \frac{49}{12m}\sum_{\mathbf{n}} a^\dagger(\mathbf{n})a(\mathbf{n})
           - \frac{3}{4m}\sum_{\mathbf{n},i} 
             \sum_{\langle \mathbf{n}\;\mathbf{n'}\rangle _i} 
             a^\dagger(\mathbf{n'})a(\mathbf{n})\\
           &+ \frac{3}{40m}\sum_{\mathbf{n},i}
             \sum_{\langle \langle \mathbf{n}\;\mathbf{n'}\rangle \rangle_i} a^\dagger(\mathbf{n'})a(\mathbf{n}) \\
           &- \frac{1}{180m}\sum_{\mathbf{n},i} 
             \sum_{\langle \langle \langle \mathbf{n}\;\mathbf{n'}\rangle \rangle \rangle _i}
             a^\dagger(\mathbf{n'})a(\mathbf{n}), 
\end{split}
\end{eqnarray}
which reproduces the free particle continuum dispersion relation up to lattice artifacts of size $O(a_{\rm latt}^6)$.
Following the notation in Ref.~\cite{Lu:2018bat}, we define the nonlocal annihilation and creation operators for real parameter $s_{NL}$ as
\begin{gather}
    a_{\sigma,NL}(\textbf{n}) = a_{\sigma}(\mathbf{n})+s_{NL}\sum_{\langle \mathbf{n} \, \mathbf{n'} \rangle }a_{\sigma}(\mathbf{n'}),\\
    a^{\dagger}_{\sigma,NL}(\mathbf{n}) = a^{\dagger}_{\sigma}(\mathbf{n})+s_{NL}\sum_{\langle \mathbf{n} \, \mathbf{n'} \rangle }a^{\dagger}_{\sigma}(\mathbf{n'}).
\end{gather}We define the point-like density operator,
\begin{equation}
    \rho(\mathbf{n}) = \sum_{\sigma}a^\dagger_{\sigma}(\mathbf{n})a_{\sigma}(\mathbf{n}),
\end{equation}
and the smeared nonlocal density,
\begin{equation}
    \rho_{NL}(\mathbf{n}) = \sum_{\sigma} a^\dagger_{\sigma,NL}(\mathbf{n})a_{\sigma,NL}(\mathbf{n}).
\end{equation}
The lattice interaction we use for the unitary Fermi gas simulations has the form
\begin{equation}
    V = \frac{C_0}{2}\sum_{\mathbf{n}}:\rho_{NL}(\mathbf{n})\rho_{NL}(\mathbf{n}):.
\end{equation}
Here the :: symbol indicates normal ordering, where all creation operators are on the left and annihilation operators are on the right.  The full Hamiltonian is then
\begin{equation}
    H = H_{\rm free} + V.
\end{equation}
We set the smearing parameter $s_{NL}$ to equal $-0.0100$, and the coefficient $C_0$ equals $-0.5573$ in lattice units. This corresponds to infinite scattering length $a_s$ and effective range of interaction $r_e \sim 0.05$ in lattice units. As noted in the main text, these values produce a vanishing $\rho^{1/3}$ coefficient for the ground state energy ratio $\xi$.  In Fig.~\ref{Fig. A.1} we show the S-wave scattering phase shifts versus relative momentum. The phase shifts at low momenta are in excellent agreement with the unitary limit, which corresponds to $\delta_0 = 90^{\circ}$.

\subsection{Euclidean time projection and auxiliary-field Monte Carlo}

Some reviews of the lattice methods presented here can be found in Ref.~\cite{Lee:2008fa,Lahde:2019npb}.  We define the transfer matrix $M$ as a normal-ordered time evolution operator over a temporal step $a_t$,
\begin{equation}
    M = :\exp[-Ha_t]:.
\end{equation}
From the expectation value of the transfer matrix, we extract the energy $E$ using the relation $\left< M \right> = e^{-Ea_t}$. Let $|\Psi_I \rangle$ be a trial wave function w nonzero overlap with the ground state. By applying powers of $M$ on $|\Psi_I \rangle$ we can project onto the ground state.  The projection amplitude for Euclidean time $t = L_ta_t$ is defined as
\begin{equation}
    Z(t) = \langle \Psi_I | M^{L_t} | \Psi_I \rangle.
\end{equation}
In general, the ground-state expectation value of any observable $\mathcal{O}$ can be calculated as $\lim_{t\rightarrow \infty} O(t)$
where
\begin{equation}
    O(t) = \frac{\langle \Psi_I|M^{L_t/2} \mathcal{O} M^{L_t/2}| \Psi_I \rangle}{\langle \Psi_I| M^{L_t} |\Psi_I \rangle}.
\end{equation}

We now decouple the particles from each other using auxiliary fields. We can rewrite the transfer matrix at time step $n_t$ as
\begin{equation}
%\begin{split}
    M^{(n_t)} = \prod_{\mathbf{n}} \left[ \sqrt{\tfrac{1}{2\pi}} \int_{-\infty}^{\infty} ds(\mathbf{n},n_t) \right] \\
    :\exp[-H_{\rm free} a_t - V_s^{(n_t)}\sqrt{a_t} - V_{ss}^{(n_t)}]:. 
    \end{equation}
where
\begin{equation}
    V_s^{(n_t)} = \sum_{\mathbf n} \sqrt{-C_0}\sum_{\mathbf{n}} s(\mathbf{n}, n_t) \rho_{NL}(\mathbf{n}),
\end{equation}
and
\begin{equation}
    V_{ss}^{(n_t)} = \frac{1}{2} \sum_{\mathbf n}  s^2({\mathbf n},n_t),
\end{equation}
where $s({\mathbf n},n_t)$ is a real-valued auxiliary variable at every lattice site. If we take the trial state $|\Psi_I \rangle$ to be a Slater determinant of single nucleon wave functions, then the auxiliary field formalism allows us to write the many-body amplitude $Z(t)$ as product of single particle amplitudes $Z_{ij}(t)$ between initial particle $j$ and final particle $i$.  After antisymmetrization of particles, we are left with the determinant of the single particle amplitude matrix $Z_{ij}(t)$.

\subsection{Euclidean time extrapolation}

Let us label the eigenstates of $H$ as $| \Psi_k \rangle $ with eigenvalues $E_k$ in nondecreasing order,
\begin{equation}
    E_0 \le E_1 \cdots \le E_k \cdots.
\end{equation}
Let $c_k$ be the overlap between our initial trial state $| \Psi_I \rangle$ and the energy eigenstate $k$, 
\begin{equation}
    c_k = \langle \Psi_k | \Psi_I \rangle,
\end{equation}
where by assumption $c_0$ is nonzero.  We can then write the amplitude $Z(t)$ in terms of its spectral decomposition,
\begin{equation}
    Z(t) = \sum_{k} |c_k|^2 e^{-E_kt},
\end{equation}
and the expectation value of the Hamiltonian has the form
\begin{equation}
    E(t) = \frac{\sum_{k} E_k |c_k|^2 e^{-E_kt}}{\sum_{k} |c_k|^2 e^{-E_kt}}.
\end{equation}
In the limit of large positive $t$, we find that
\begin{equation} \label{Eq. (22)}
    E(t) \approx E_0 + |c_1|^2/|c_0|^2 \Delta E e^{-\Delta E t},
\end{equation}
where $\Delta E = E_1 - E_0$ is the energy gap in the spectrum. 

For an operator that does not commute with the Hamiltonian, the time dependence of the expectation value is
\begin{equation}
    O(t) = \frac{\sum_{k,k'} c_k^*O_{k,k'} c_{k'} e^{-E_kt/2}e^{-E_k't/2}}{\sum_{k} |c_k|^2 e^{-E_kt}}.
\end{equation}
where $O_{k,k'}=\left<\Psi_k| O |\Psi_{k'}\right>$.  In this case behavior at large $t$ will have the form
\begin{equation}
    E(t) \approx O_{0,0} + 2\,{\rm Re}(O_{0,1}c_1/c_0)e^{-\Delta E t/2},
    \label{Et/2}
\end{equation}
where $\Delta E = E_1 - E_0$ is again the energy gap.

\section{Results for ground state energy}
We perform lattice simulations on a periodic cubic with length $L$ in lattice units ranging from $L = 5$ to $L = 11$.  We use the Euclidean-time evolution operator $\exp(-Ht)$ to project the ground state from the initial state $| \Psi_I \rangle $. $| \Psi_I \rangle $ is composed of the ground state of a free Fermi gas with $N_{\uparrow}=N_{\downarrow}=33$, for a total of $N=66$ particles.  As noted in Ref.~\cite{Bour:2011xt}, there are two reasonable definitions of the energy ratio $\xi$ for finite $N_{\uparrow},N_{\downarrow}$.  One is to simply take the ratio of the interacting and non-interacting systems for the same number of particles, $N_{\uparrow}$ and $N_{\downarrow}$, and the same volume, $L^3$.  We denote this ratio of finite system energies as $\xi^{\rm finite}_{N_{\uparrow},N_{\downarrow}}$.  But one can also take the thermodynamic limit for the free Fermi gas energy, $E_{FG} = 3Nk_{F}^2/(10m)$, where the Fermi momentum is $k_{F} = (3\pi^2 N)^{1/3}\hbar/L$ for the spin-balanced system $N_\uparrow = N_\downarrow = N/2$.  We denote this ratio as $\xi^{\rm thermo}_{N_{\uparrow},N_{\downarrow}}$.  Both $\xi_{N_{\uparrow},N_{\downarrow}}$ and $\xi^{\rm thermo}_{N_{\uparrow},N_{\downarrow}}$ become $\xi$ in the thermodynamic limit and the difference between the two gives a rough estimate of finite-size errors.

\begin{figure}[h!]
    \centering
    \includegraphics[width=0.9\textwidth]{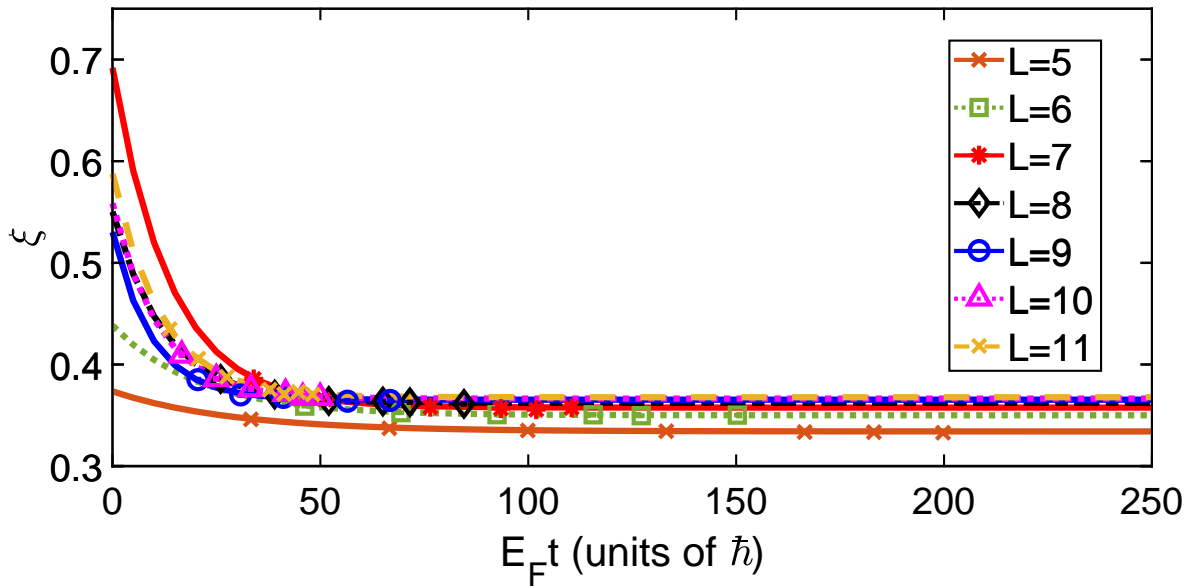}
    \captionsetup{justification=raggedright}
    \caption{Plot of $\xi^{\rm thermo}_{33,33}(t)$ versus $E_F t$ for volumes $L=5$ to $L=11$.  We also show the fitted exponential curves used to extrapolate to infinite $t$.\\}
    \label{Fig. A.2}
\end{figure}

\begin{figure}[h!]
    \centering
    \includegraphics[width=0.9\textwidth]{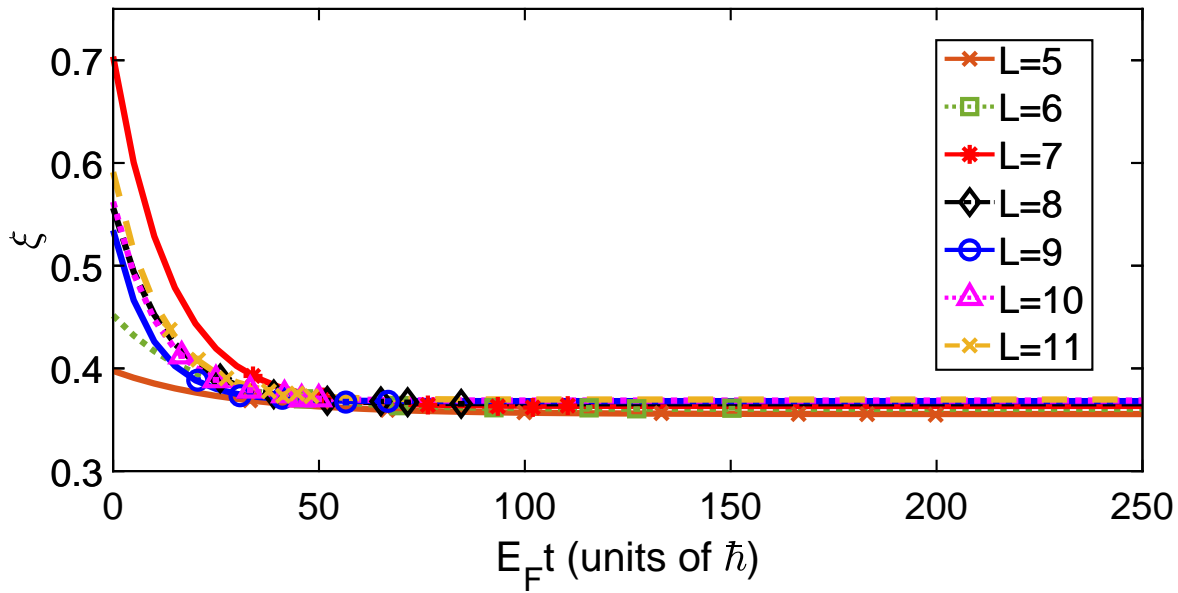}
    \captionsetup{justification=raggedright}
    \caption{Plot of $\xi^{\rm finite}_{33,33}(t)$ versus $E_F t$ for volumes $L=5$ to $L=11$.  We also show the fitted exponential curves used to extrapolate to infinite $t$.\\}
    \label{Fig. A.3}
\end{figure}

 In Fig.~\ref{Fig. A.2} we show a plot of $\xi^{\rm thermo}_{33,33}(t)$ versus $E_F t$ for volumes $L=5$ to $L=11$.  We also show the fitted exponential curves used to extrapolate to infinite $t$.  Similarly, in Fig.~\ref{Fig. A.3} we show a plot of $\xi^{\rm finite}_{33,33}(t)$ versus $E_F t$ for volumes $L=5$ to $L=11$ and the fitted exponential curves used to extrapolate to infinite $t$.

\begin{figure}[t!]
    \centering
    \includegraphics[width=0.9\textwidth]{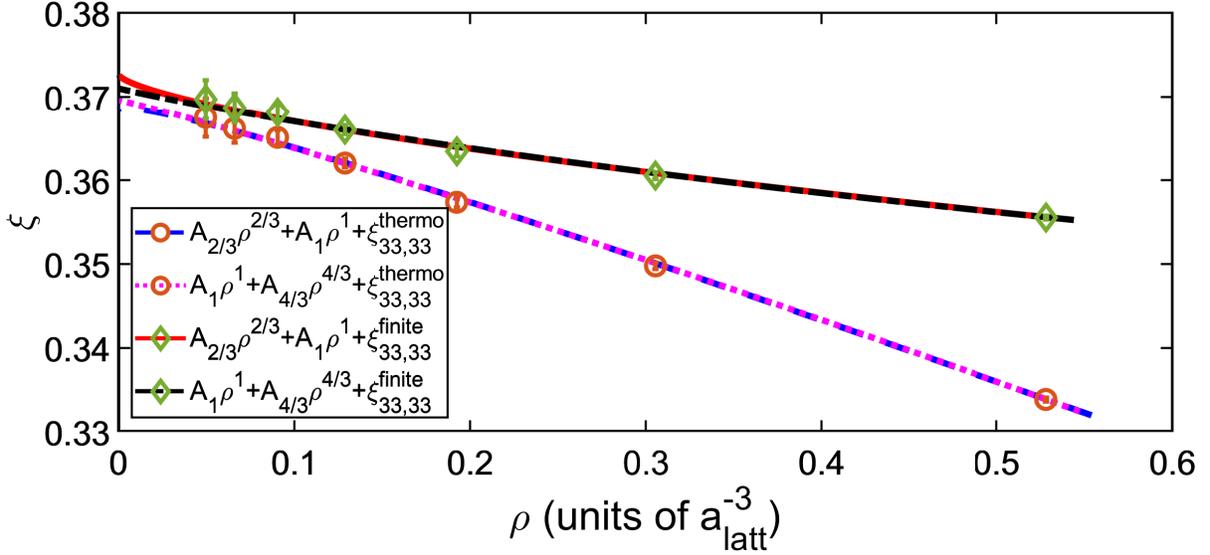}
    \captionsetup{justification=raggedright}
    \caption{$\xi_{33,33}^{\rm thermo}$ and  $\xi^{\rm finite}_{33,33}$ for lattice sizes $L = 5$ to $11$ and $N_\uparrow = N_\downarrow = 33$ particles versus particle density $\rho$. }
    \label{Fig. 1}
\end{figure}

In Fig.~\ref{Fig. 1}, we show results for $\xi^{\rm finite}_{33,33}$ and $\xi^{\rm thermo}_{33,33}$ as a function of the particle density $\rho = N(a_{\rm latt}L)^{-3}$. For each case we expand in powers of $L^{-1}$ or $\rho^{1/3}$ to extrapolate to the dilute limit.  The range of the nonlocal S-wave interaction was adjusted so that the leading order $\rho^{1/3}$ dependence vanishes.  As expected from theoretical considerations, this occurs at or very close to the point where the S-wave effective range parameter vanishes.  It turns out that for our chosen lattice action, the coefficient of $\rho^{2/3}$ is also very small.  This allows for an accurate extrapolation to the zero density limit.  We perform two fits for the residual $L$ dependence, one using the form $A_{2/3}\rho^{2/3} + A_1\rho^1 + \xi$ and the other using $A_1\rho^1 + A_{4/3}\rho^{4/3} + \xi$.   Model averaging over the two fits, we obtain the values $\xi^{\rm thermo}_{33,33} = 0.369(2)$ and $\xi^{\rm finite}_{33,33} = 0.372(2)$.  This is consistent with the results $\xi^{\rm thermo}_{33,33} = 0.374(5), 0.372(3), 0.375(5)$ obtained in Ref.~\cite{paper 4.32} using three different lattice actions.  This provides a good benchmark test for our new lattice calculations.  While we have not performed a extrapolation to the thermodynamical limit, $N\rightarrow \infty$, the results of Ref.~\cite{paper 4.32} and other studies have found that the ground state energy for the $N=66$ system is less than one percent away from the thermodynamic limit.

Our results agree with the results in Ref.~\cite{paper 4.32, paper 4.32+41, paper 4.33} as well as the values $\xi = 0.366_{-0.011}^{+0.016}$ by the lattice Monte Carlo calculations \cite{paper 4.38+1} and the value $\xi=0.367(7)$ obtained by a zero-temperature extrapolation from the finite temperature simulations \cite{paper 4.38+2} and the upper bound 0.383(1) calculated by fixed-node diffusion Monte Carlo combined with density functional theory \cite{paper 4.32+41, paper 4.33}. Our results also agree with the experimental value 0.376(4) \cite{paper 4.38}.  Further details of the calculations are provided in the Supplemental Materials.

%%%%%%%%%%%%%%%%%% NEW

\section{Results for the condensate fraction}

\begin{figure}[t!]
    \centering
    \includegraphics[width=0.9\textwidth,keepaspectratio]{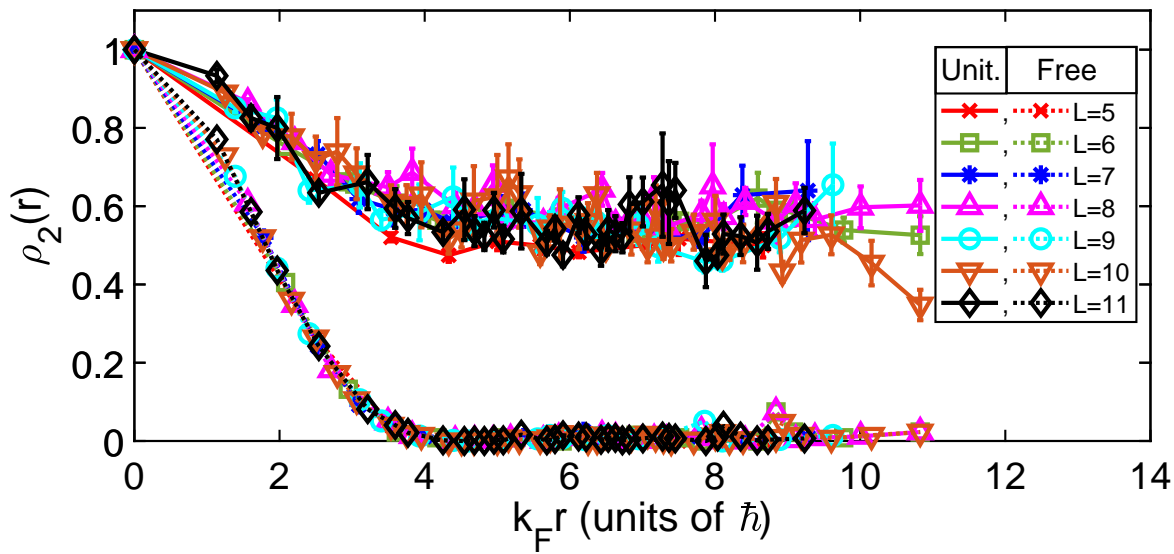}
    \captionsetup{justification=raggedright}
    \caption{Pair-pair correlation functions for unitary Fermi gas (solid curves) and free Fermi gas (dashed curves) for $N=66$ particles with volumes $L = 5$ to $L=11$.} 
    \label{Fig. 2}
\end{figure}

Having verified the ground state energy benchmark, we now perform calculations of the superfluid condensate and pairing wave function.   For these calculations we compute one- and two-body density matrices in the ground state.  We first establish that the ground state is a S-wave superfluid.  For this we compute the pair-pair correlation function, which is simply the TBDM for the case where $\mathbf{r'_1}=\mathbf{r'_2}$ and $\mathbf{r_1}=\mathbf{r_2}$.  In order to work with universal quantities that are independent of short distance physics, we divide the pair-pair correlation function by its value at zero distance \cite{Lee:2006vp},
\begin{equation}
    \rho_2(\mathbf{r}) = \langle a^{\dagger}_\uparrow(\mathbf{r})a^{\dagger}_\downarrow(\mathbf{r}) a_\downarrow(\vec{0})
    a_\uparrow(\vec{0}) \rangle / \Gamma ,
\end{equation}
where $\Gamma$$\,=\,$$\langle a^{\dagger}_\uparrow(\vec{0})a^{\dagger}_\downarrow(\vec{0}) a_\downarrow(\vec{0}) a_\uparrow(\vec{0})  \rangle$.  In Fig.~\ref{Fig. 2}, we show results for  $\rho_2(r)$  versus $k_Fr$ for the unitary Fermi gas as well as the free Fermi gas for $N=66$ particles with volumes $L = 5$ to $L=11$.  In each case we see universal behavior that is independent of $L$. There is no long-range order for free gas, but the unitary gas has off-diagonal long-range order consistent with an S-wave superfluid.

\begin{figure}[t!]
    \centering
    \includegraphics[width=0.9\textwidth,keepaspectratio]{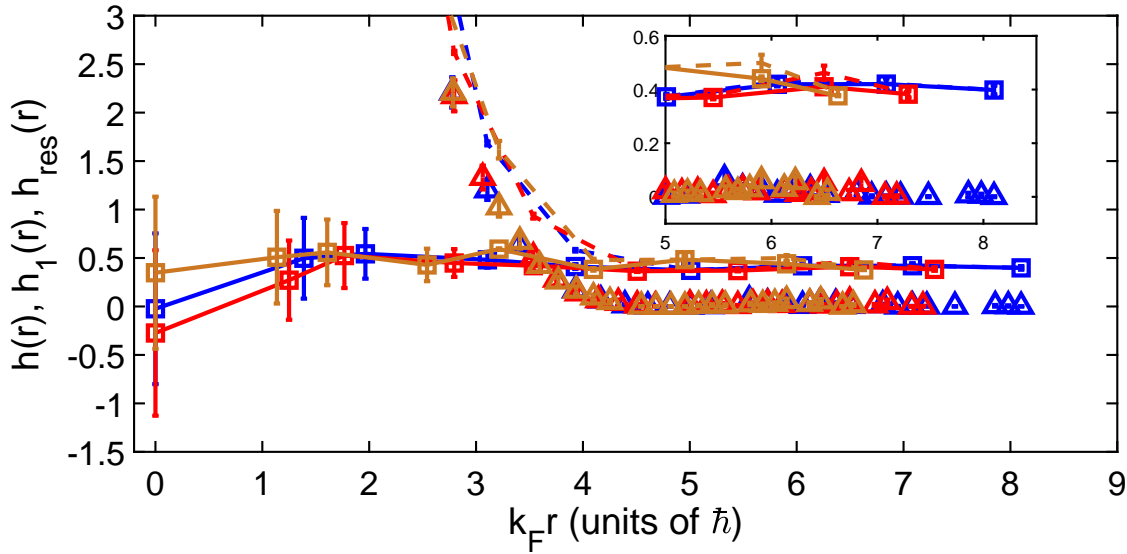}
    \captionsetup{justification=raggedright}
    \caption{Lattice results for the project TBDM $h(r)$ (circles and dashed lines), $h_1(r)$ (triangles) and $h_{\rm res}(r)$ (squares and solid lines), for $N=66$ particles with lattice sizes $L = 9$ in blue, $L = 10$ in red, and $L = 11$ in ochre.}
    \label{Fig. 3}
\end{figure}

We now calculate the condensate fraction $\alpha$. Let us consider the projected TBDM \cite{paper 4.45, paper 4.46},
\begin{equation}
    h(\mathbf{r}) = \frac{2}{N}\int \rho_2(\mathbf{r_1+r},\mathbf{r_2+r},\mathbf{r_1},\mathbf{r_2}) d\mathbf{r_1} d\mathbf{r_2}, \label{hr}
\end{equation}
Combining Eq.~\ref{TBDM} and Eq.~\ref{hr}, we find that $\alpha$ equals $h(\mathbf{r})$ in the limit of large $|\mathbf{r}|$. For small $|\mathbf{r}|=0$, however, $h(\mathbf{r})$ also contains contributions from the product of OBDMs for each of the two particles.  In order to speed the convergence to the asymptotic limit, we define the residual of the projected TBDM as
    \begin{align}
        h_{\rm res}(\mathbf{r})
        & = h(\mathbf{r}) - h_1(\mathbf{r}), \\
      h_1(\mathbf{r}) & = \frac{2}{N}[\rho_1(\mathbf{r})L^3]^2,
    \end{align}
where $\rho_1(\mathbf{r})=\rho_{1,\sigma}(\mathbf{r},\mathbf{0})$.  In Fig. \ref{Fig. 3}, we show the results for $h(r)$, $h_1(r)$, and $h_{\rm res}(r)$ versus $k_F r$ for volumes $L = 9$ to $L=11$. The asymptotic behavior for $h(r)$ at large $r$ reveals the value of the condensate fraction $\alpha$. Although $(2/N)[\rho_1(\mathbf{r})L^3]^2$ vanishes at a large $r$, its contribution at a small $r$ cannot be neglected. By subtracting its contribution from $h(r)$, we see that $h_{\rm res}(r)$ yields a good estimate for $\alpha$ over a wide range of values $k_F r$.

%%%%%%%%%%%%%%%

\begin{figure}[h!]
    \centering
    \includegraphics[width=0.9\textwidth,keepaspectratio]{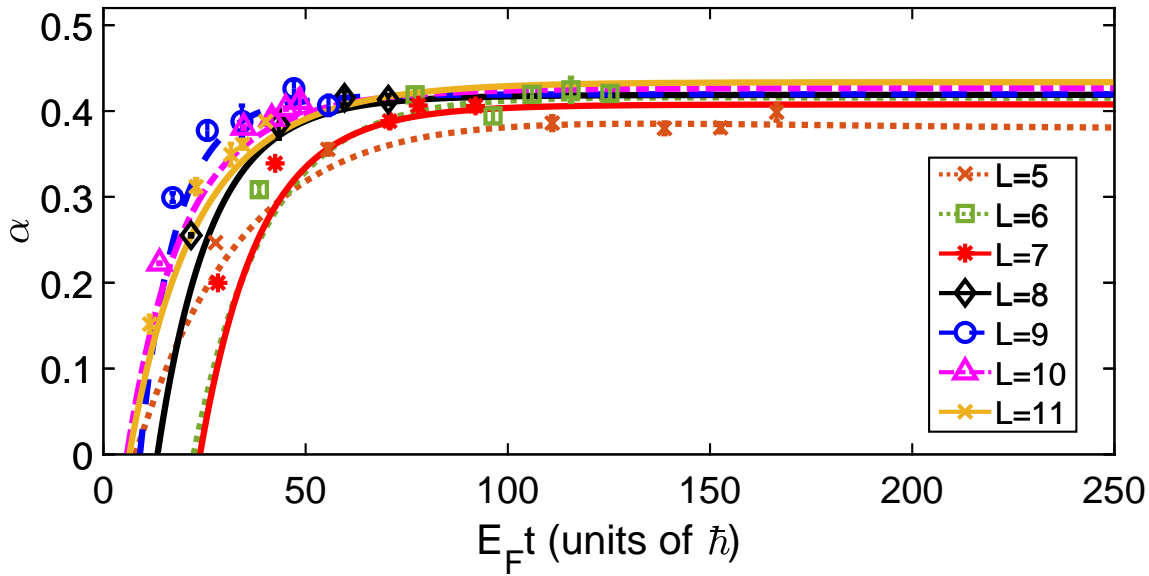}
    \captionsetup{justification=raggedright}
    \caption{Condensate fraction $\alpha$ versus $E_F t$ for $N=66$ particles  with volumes $L=5$ to $L=11$.  We also show the fitted curves used to extrapolate to infinite Euclidean time $t$.} 
    \label{Fig. A.4} 
\end{figure}

We compute the condensate fraction $\alpha$ by calculating the average value of $h_{\rm res}(\mathbf{r})$ over the entire spatial lattice.  When we compute this lattice average for different projection times, we obtain the results shown in Fig.~\ref{Fig. A.4} for $N=66$ particles with volumes $L=5$ to $L=11$.  For the extrapolation to infinite Euclidean time, we use the functional form described in Eq.~(\ref{Et/2}), and these are also shown in Fig.~\ref{Fig. A.4}.
 
%%%%%%%%%%%%%%%

\begin{figure}[t!]
    \centering
    \includegraphics[width=0.9\textwidth,keepaspectratio]{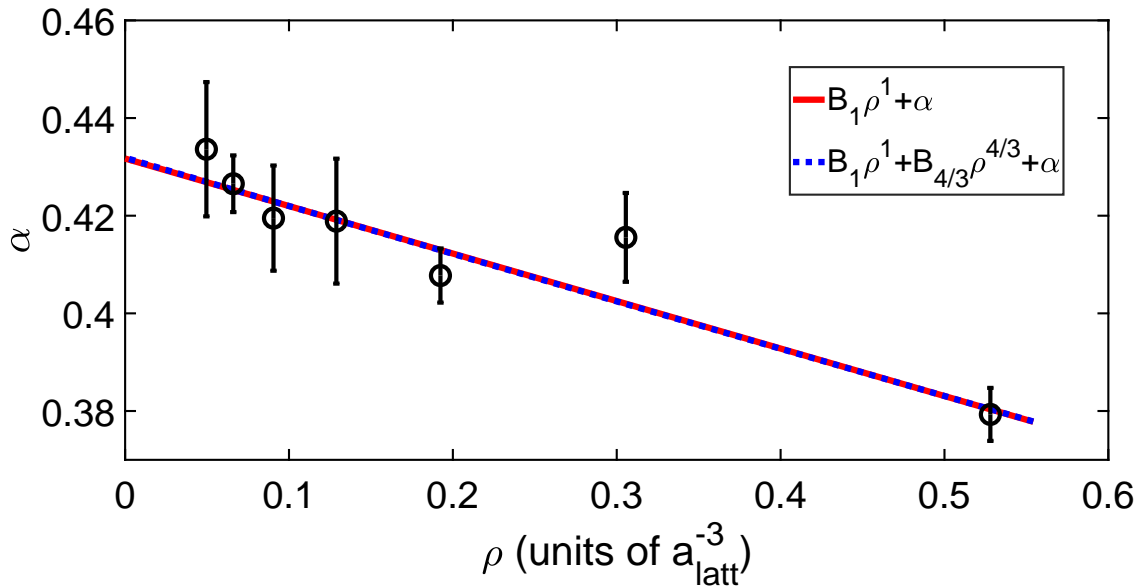}
    \captionsetup{justification=raggedright}
    \caption{Condensate fraction $\alpha$ versus density $\rho$ for $N=66$ particles with volumes from $L = 5$ to $L = 11$.} 
    \label{Fig. 4} 
\end{figure}

In Fig.~\ref{Fig. 4}, we summarize our results for the condensate fraction $\alpha$ versus the particle density $\rho$, along with fit functions of the form $B_1\rho^{1} + B_{4/3}\rho^{4/3}+\alpha$ and $B_1\rho^1 +\alpha$ for the residual density dependence.
Model averaging over the two fits, we find that condensate fraction $\alpha$ at unitarity is 0.43(2) for the $N=66$ system.  Our estimates agree with the experimental values $0.46(7)$ \cite{paper 4.48} and $0.47(7)$ \cite{Roati} and the value 0.43 predicted in Ref.~\cite{paper 4.48+2}. Our value is also consistent with zero temperature extrapolations of previous lattice calculations which have large uncertainties but favor a range between 0.4 to 0.5 \cite{Bulgac:2008,Jensen:2018opr}.

\section{Results for the pairing wave function}

\begin{figure}[h!]
    \centering
    \includegraphics[width=0.9\textwidth,keepaspectratio]{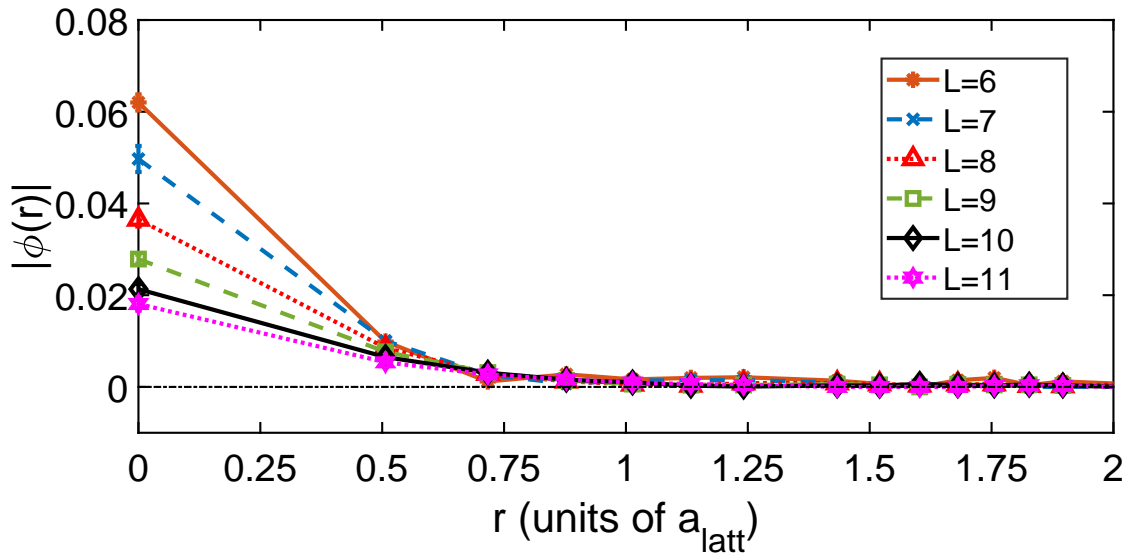}
    \captionsetup{justification=raggedright}
    \caption{Pair wave function, $|\phi(r)|$, for $N=66$ particles with volumes $L=6$ to $L=11$.} 
    \label{Fig. A.8} 
\end{figure}

We have also calculated the pair wave function $\phi(|\mathbf{r}|)$ defined in Eq.~\ref{TBDM}. For the pair wave functions, we find that extrapolation of the pair wave function to infinite $t$ introduces more error than simply taking the pair wave function in the asymptotic region where the exponential dependence on $t$ is negligible.  In Fig. \ref{Fig. A.8}, we show the pair wave functions $|\phi(r)|$ for $N=66$ particles with volumes $L=6$ to $L=11$.
Since the pairing interaction is in the S-wave channel, we fit the tail of the pair wave function to the asymptotic form expected for a bound state with a finite range S-wave interaction, $A\exp(-r/\zeta_p)/r$, where $\zeta_p$ is the pair correlation length, and $A$ is a normalization factor. 

\begin{figure}[t!]
    \centering
    \includegraphics[width=0.9\textwidth,keepaspectratio]{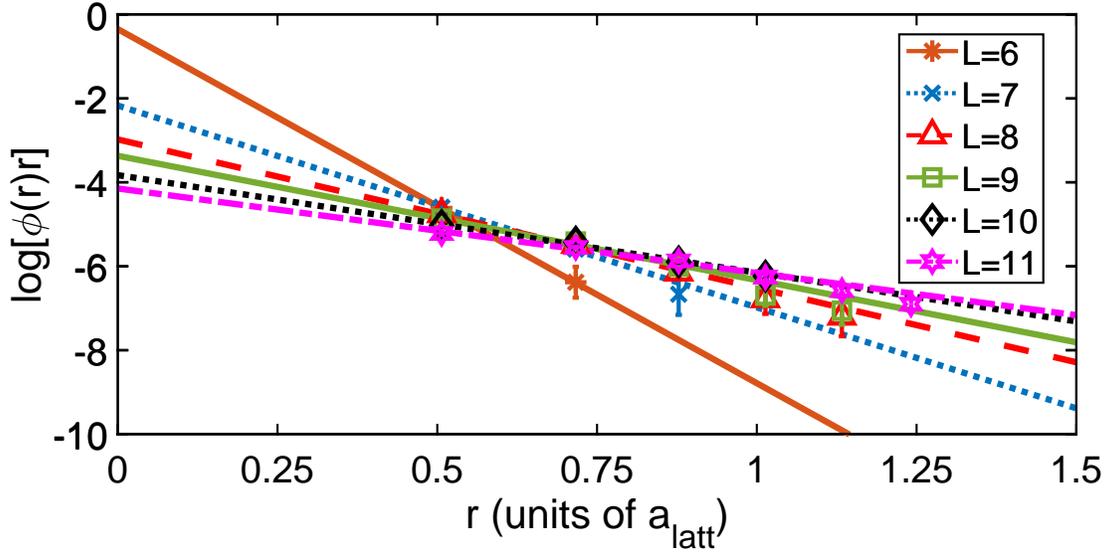}
    \captionsetup{justification=raggedright}
    \caption{Plot of $\log[\phi(r)r]$ for $N=66$ particles with volumes from $L=6$ to $L=11$.  The slope at large $r$ corresponds to $\zeta_p^{-1}$. \\} 
    \label{Fig. 5} 
\end{figure}

\begin{figure}[t!]
    \centering
    \includegraphics[width=0.9\textwidth,keepaspectratio]{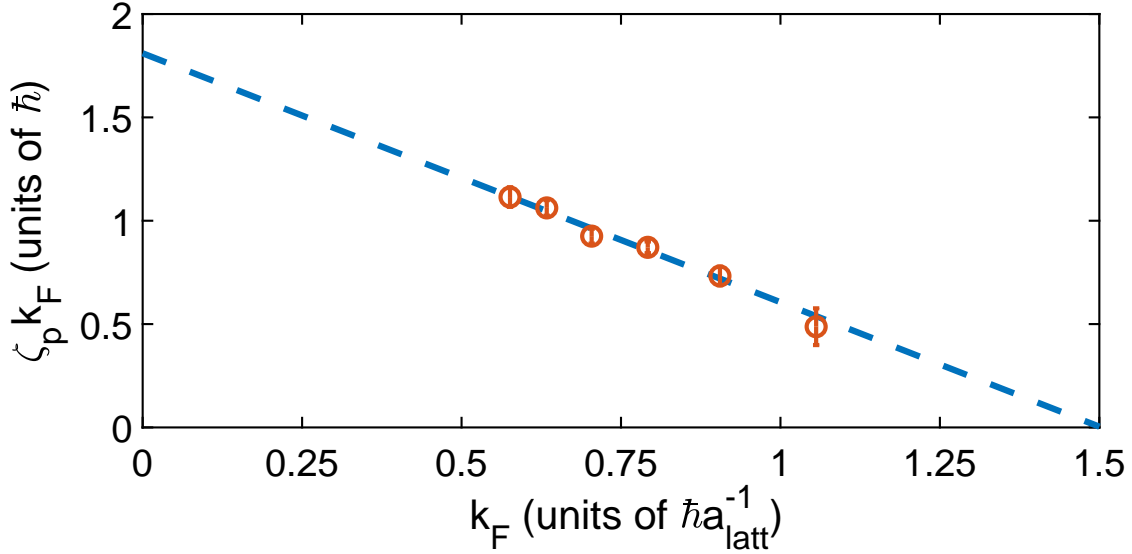}
    \captionsetup{justification=raggedright}
    \caption{Plot of $\zeta_pk_F$ as a function of $k_F$ for $N=66$ particles with volumes from $L=6$ to $L=11$. We also show the fit function $C_{1/3}\rho^{1/3} + \zeta_pk_F$. \\} 
    \label{Fig. 6} 
\end{figure}

In Fig.~\ref{Fig. 5} we plot $\log[\phi(r)r]$ versus $r$ for $N=66$ particles with volumes from $L=6$ to $L=11$.  The slope of the $\log[\phi(r)r]$ plot corresponds to $-\zeta_p^{-1}$.  In Fig.~\ref{Fig. 6} we show the results obtained for $\zeta_pk_F$ versus $k_F$ for  $L=6$ to $L=11$.  In order to extrapolate to the dilute limit we use the fit function $C_{1/3}\rho^{1/3} + \zeta_pk_F$.  The data points and fit function are shown in Fig. \ref{Fig. 6}. We find that the pair correlation length is $\zeta_p k_F = 1.8(3) \hbar$. Ideally one would like to have more data at smaller lattice spacings in order to pin down the continuum limit extrapolation with more accuracy.  We hope to explore this in future calculations.

To our knowledge this is the first {\it ab initio} calculation of the pair correlation length in the unitary limit.  We can compare this result with the corresponding result from BCS theory, $\zeta_p k_F = 2\hbar/(\pi \delta)$ where $\delta$ is the ratio of the pairing gap to $E_F$ \cite{Annett}.  If one carries over the same formula to the unitary limit, the estimate is $\zeta_p k_F \sim 1.3 \hbar$, which is not too far from our calculated result.

\section{Summary and conclusions}
We have performed auxiliary-field lattice Monte Carlo simulations of the unitary Fermi gas using a novel lattice interaction that accelerates the approach to the continuum limit and allows for an accurate estimate of the two-body density matrix for 66 particles.  We computed the ground state energy of 33 spin-up and 33 spin-down particles and found good agreement with the latest theoretical and experimental determinations.  We found $E_0/E_{FG}= 0.369(2), 0.372(2)$ using two different definitions of the finite-system energy ratio.  We then determine the condensate fraction by measuring off-diagonal long-range order in the two-body density matrix and found that the fraction of condensed pairs is $\alpha = 0.43(2)$.  We then extracted the pairing wave function and found the pair correlation length to be $\zeta_pk_F = 1.8(3) \hbar$.  Our results for the condensate fraction and pair correlation length are free from the systematic errors associated with one-sided estimates of the two-body density matrix.  By using an interaction with a simpler extrapolation to the continuum limit, we were able to obtain more accurate results in the zero temperature limit than previous lattice calculations.

The methods presented here have general applicability and can be applied to other S-wave superfluid systems.  In particular, our lattice approach can be used to investigate the superfluid properties of neutron gases, which is essential for understanding the structure and evolution of neutron stars.  So long as the problem of Monte Carlo sign oscillations is under control, one can also employ auxiliary-field lattice simulations to study more exotic systems such as P-wave and D-wave superfluids.

\paragraph*{Acknowledgments}

We are grateful for discussions with Witek Nazarewicz.  We acknowledge partial financial support from the U.S. Department of Energy (DE-SC0018638 and DE-AC52-06NA25396). The computational resources were provided by the Julich Supercomputing Centre at Forschungszentrum J\"ulich, Oak Ridge Leadership Computing Facility, RWTH Aachen, and Michigan
State University.

\end{document}